\documentclass[12pt,preprint]{emulateapj}

\shorttitle{Model for Mira's comet}
\shortauthors{Raga et al.}

\begin{document}

\slugcomment{Accepted for publication in ApJL}

\title{A latitude-dependent wind model for Mira's cometary head}

\author{A. C. Raga\altaffilmark{1}, J. Cant\'o\altaffilmark{2},
F. De Colle\altaffilmark{3},
A. Esquivel\altaffilmark{1},
P. Kajdic\altaffilmark{4},
A. Rodr\'\i guez-Gonz\'alez\altaffilmark{1}, 
P. F. Vel\'azquez\altaffilmark{1}}
\email{raga@nucleares.unam.mx}

\altaffiltext{1}{Instituto de Ciencias Nucleares, Universidad
Nacional Aut\'onoma de M\'exico, Ap. 70-543,
04510 D.F., M\'exico}
\altaffiltext{2}{Instituto de Astronom\'\i a, Universidad
Nacional Aut\'onoma de M\'exico, Ap. 70-543,
04510 D.F., M\'exico}
\altaffiltext{3}{Dublin Institute for Advanced Studies (DIAS),
31 Fitzwilliam Place, Dublin 2, Ireland}
\altaffiltext{4}{Instituto de Geof\'\i sica, Universidad
Nacional Aut\'onoma de M\'exico, 04510 D.F., M\'exico}

\begin{abstract}
We present a 3D numerical simulation of the recently
discovered cometary structure produced as Mira travels through
the galactic ISM. In our simulation, we consider that Mira ejects
a steady, latitude-dependent wind, which interacts with a homogeneous,
streaming environment. The axisymmetry of the problem is broken
by the lack of alignment between the direction of the relative motion
of the environment and the polar axis of the latitude-dependent
wind. With this model, we are able to produce a cometary head
with a ``double bow shock'' which agrees well with the structure
of the head of Mira's comet. We therefore conclude that a
time-dependence in the ejected wind is not required for
reproducing the observed double bow shock.
\end{abstract}


\section{Introduction}

Martin et al. (2007) describe UV observations of a cometary
structure extending over $2^\circ$ in the sky, with its head
centered on Mira. These observations were carried out with
the GALEX satellite, with a filter with a central
wavelength $\lambda_c=1516$~\AA\ and a FWHM $\Delta \lambda
=256$~\AA . Martin et al. (2007) speculate that part of the
observed emission could be associated with the fluorescent
UV lines of the H$_2$ molecule. Also, part of the emission
(particularly in the head of the cometary structure) could
correspond to emission from ionized species (e. g., from
C~IV) with lines which fall in the bandwidth of the filter.
We present a schematic diagram of the UV image of Mira's
comet of Martin et al. (2007) in Figure 1.

Mira is a binary system with an AGB star and a less
luminous companion which is probably a white dwarf.
The binary system moves through the plane of the galaxy
with a spatial velocity of $\approx 130$~km~s$^{-1}$,
at an orientation of $\approx 30^\circ$ with respect to the plane
of the sky, as can be deduced from the proper motion
(Turon et al. 1993), radial velocity (Evans 1967)
and Hipparcos-based distance (107 pc, Knapp et al.
2003). The proper motion is directed at an angle
of $187^\circ$ (measured E from N, see Wareing et al. 2007).

Wareing et al. (2007) computed a 3D simulation, in which
they model this object as the interaction between an isotropic
wind from the asymptotic giant branch (AGB)
primary star and a streaming, homogeneous medium.
Their simulation produces structures which are similar to the
observed cometary structure. However, it is clear that the
predicted flow does not show the detailed structure of condensations
observed in the head of Mira's comet (see Figure 3 of Martin et al.
2007).

In the present paper we explore whether or not the ``double bow
shock'' structure of the head of Mira's comet can be reproduced
by a somewhat more complex model. In modelling the structures of
planetary nebulae (PNe), it is quite common to consider that
the wind from AGB stars has a strong, equatorial density
enhancement. Such a latitude-dependent wind can be used for
reproducing the structures of many observed PNe and Proto-PNe
(see, e.g., Balick \& Frank 2002).

In the case of Mira, we could also be seeing the interaction
of a latitude-dependent wind with an environment which sweeps past
the source. We have, however, few constraints of the characteristics and
orientation of this latitude-dependent wind. The only real observational
constraint is that the Mira binary system has an orbital plane that
lies within $\sim 30^\circ$ from
the line of sight, and at an angle of $130\to 180^\circ$
(measured E from N), with very high uncertainties (see Prieur et al.
2002). If we assume that the polar axis of the latitude-dependent
wind lies perpendicular to the orbital plane, we would then have to
place it close to the plane of the sky, and at an angle of $40\to 90^\circ$
(measured E from N). This direction for the polar axis, taken together
with the direction of Mira's proper motion (see above) implies that
the polar axis of the outflow has to lie at an angle
$\alpha\sim 33\to 83^\circ$ with respect to the direction of
relative motion of the ISM which is streaming past Mira's wind.

After exploring a limited set of $\sim 10$ parameter combinations,
we have chosen a model in which the angle between the polar axis of
the latitude-dependent wind and the direction of the streaming
environment has a value of $70^\circ$, which is consistent
with the observations described above. In \S 2, we describe
the choice of parameters for our model, and the results from
the numerical time-integration. In \S 3, we discuss the results.

\section{The model}

\begin{figure}
\epsscale{0.9}
\plotone{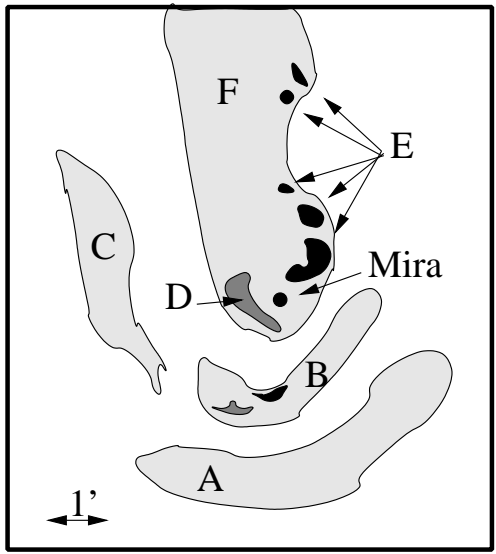}
\caption{Schematic diagram of the UV image of the head of
Mira's comet, showing the structures seen in Figure 3 of Martin
et al. (2007). The proper motion of Mira is directed at an angle
of $\approx 7^\circ$ to the W of S (N is up and E is left),
approximately aligned with the large scale structure of the cometary
tail. Therefore, when comparing this diagram with the model predictions
(Figures 2-4), one has to allow for this small deviation from the
vertical direction of Mira's motion. In this Figure, we introduce
labels A-F in order to identify the different structures which are
discussed in the text.}
\label{fig1}
\end{figure}

We model the cometary tail of Mira as the interaction of a
latitude-dependent wind with a homogeneous environment in relative
motion with respect to Mira. For the stellar wind, we consider
a density of the form~:
\begin{equation}
\rho(r,\theta)={A\over r^2} f(\theta)\,,\,\,\,\,\,
f(\theta)=\xi-\left(\xi-1\right)\,|\cos\theta|^p\,,
\label{rho}
\end{equation}
where $r$ is the spherical radius and $\theta$ is the
polar angle (measured
out from the symmetry axis of the wind). In this equation, $A$ is a
scaling constant, $\xi>1$ is the equator-to-pole
density ratio, and $p$ is a constant that determines the degree
of flattening towards the equator of the density stratification.
We have taken this form of the anisotropy function $f(\theta)$
from Riera et al. (2005).

We now consider a latitude-dependent wind velocity of the form
\begin{equation}
v(\theta)={v_0\over \sqrt{f(\theta)}}\,.
\label{v}
\end{equation}
where $v_0$ is the velocity of the wind in the polar direction.
With this choice, the ram pressure $\rho v^2$ of the wind is
isotropic.

One can relate the scaling constant $A$ with the mass loss rate
of the wind ${\dot M}_w$ through the relation
\begin{eqnarray}
{\dot M}_w=4\pi r^2\int_0^{\pi/2}\rho(r,\theta)v(r,\theta)
\sin\theta\,d\theta= \nonumber \\ 
={4\pi A v_0} \int_0^{\pi/2} \left[f(\theta)\right]^{1/2}
\sin\theta\,d\theta\,.
\label{mint}
\end{eqnarray}
We now choose an anisotropy function with $p=1/2$. With this
choice, one can compute analytically the integral in equation
(\ref{mint}) to obtain~:
\begin{equation}
{\dot M}_w=\left({4\pi A v_0}\right)
{{8\xi^{5/2}-20\xi+12}\over {\left(\xi-1\right)^2}}\,.
\label{m2int}
\end{equation}
We now choose a $\xi=20$ equator to pole density ratio, for which
through equation \ref{m2int} we obtain $A=0.390 {\dot M}_w/(4\pi v_0)$.

\begin{figure}
\epsscale{1.1}
\plotone{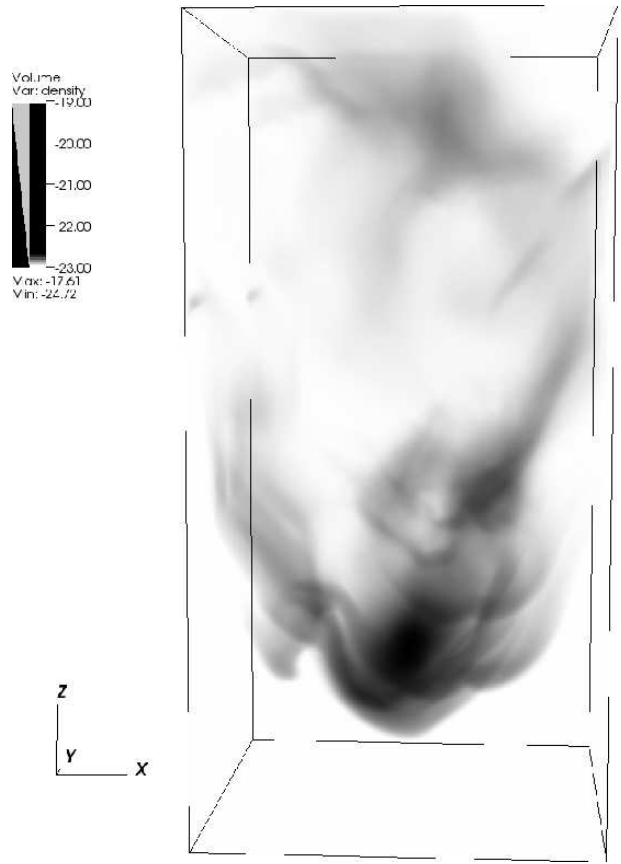}
\caption{Volume rendition of the density structure obtained for
a $t=8\times 10^4$~yr time integration from the simulation described in the
text. The horizontal extent of the displayed domain is of
$1.5\times 10^{18}$~cm.}
\label{fig2}
\end{figure}

To complete the parameters for our anisotropic wind we
choose a polar velocity $v_0=10$~km~s$^{-1}$ and a mass loss
rate ${\dot M}_w=7.7\times 10^{-7}$M$_\odot$yr$^{-1}$. We set
the wind temperature equal to $10^3$~K at a distance $r_w={8\times
10^{16}}$~cm from the star, and impose the wind at all times
within this radius.
For the environment, we assume that it enters the grid along the
$z$-axis, with a constant velocity $v_*=130$~km~s$^{-1}$, density
$n_{env}=0.1$~cm$^{-3}$ and temperature $T_{env}=10^3$~K.

Finally, we tilt the polar axis of the wind (out from which the angle
$\theta$ is measured, see equations \ref{rho} and \ref{v}) so that it
lies on an $xz$-plane, pointing at an angle $\alpha=70^\circ$ measured
from the $z$-axis. This angle is consistent with the orientation of
the orbital axis of the Mira binary system (see \S 1).

We now carry out a 3D numerical simulation with the
yguaz\'u-a code in a $(1.5,1.5,3.0)
\times 10^{18}$~cm domain. The stellar wind source is centered
on an $xy$-plane, and placed at a $z_0=6\times 10^{17}$~cm distance
from the edge of the grid along the $z$-axis. This domain is
resolved with a 5-level, binary adaptive grid giving $128\times 128
\times 256$ points at the maximum resolution. The yguaz\'u-a code
has been described in detail by Raga et al. (2000), and has been
tested with laboratory experiments and employed for computing many
different astrophysical flows (see, e.g., the review of Raga et al.
2006). We have used the version of the code in which the gasdynamic
equations are integrated together with a rate equation for the
ionization of H, and uses a parametrized cooling
function (computed as a function of the density, temperature
and H ionization fraction), which is described by Raga \& Reipurth
(2004).

The numerical simulation is started with the wind (see equations
\ref{rho}-\ref{v}) occupying all of the computational domain, and
the streaming environment entering from the bottom of the domain.
Figure 2 shows a volume rendition of the density of the flow after
a $t=8.0\times 10^4$~yr time-integration. The density stratification
shows a leading bow shock with a complex structure, and a tail with
dense clumps and filaments. Figure 3 shows the column densities
obtained from this stratification, calculated by integrating the
density along the $y$- and the $x$-axes, and also assuming
that the $z$-axis is at a $\phi=30^\circ$ angle with respect
to the plane of the sky (with the $x$-axis on this plane).
This orientation angle corresponds to the orientation of
the motion of Mira with respect to the plane of the sky (see
\S 1). The column density maps show structures with
large differences depending on the direction of the integration.

\begin{figure}
\epsscale{1.0}
\plotone{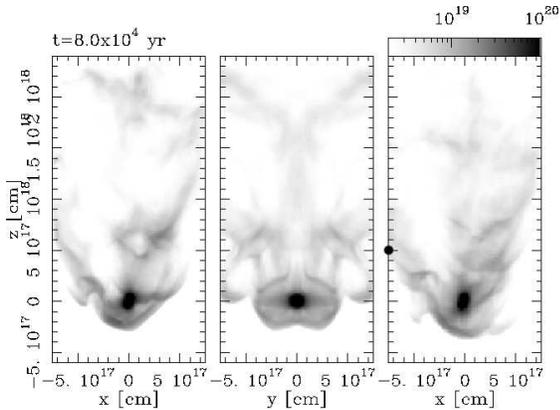}
\caption{Column density maps for $t=8\times 10^4$~yr obtained by
integrating the atom+ion column density along the $y$-axis (left)
and along the $x$-axis (center). On the right we show the column
density map for an orientation with the $x$-axis on the plane of
the sky, and the $z$-axis at an angle $\phi=30^\circ$ with respect
to this plane. The column densities (in cm$^{-2}$) are shown with
the logarithmic greyscale given by the bar on the top right. The
origin of the coordinate system coincides with the (projected)
position of the stellar wind source.}
\label{fig3}
\end{figure}

In Figure 4, we show part of the time-evolution of the column
density maps (calculated assuming a $\phi=30^\circ$ angle
between the $z$-axis and the plane of the sky)
of the region around the head of the cometary structure.
In this time-sequence, we see that the region upwind from
the star (i. e., below the star in the frames of Figure 4)
at all times has a leading thick, dense region, which
corresponds to the snowplow structure of the wind/streaming
ISM interaction. The projection of the 3D structure on
the plane of the sky has a complex shape, at some times
giving a ``double bow shock'' morphology (see, e.g., the
$t=7.0\times 10^4$~yr frame of Figure 4). This leading
bow shock has a general side-to-side asymmetry that resembles
the observations of Mira (see Figure 1).

The separation between the star and the leading shock
changes with time, giving values in the
$(3.6\to 4.2)\times 10^{17}$~cm range (measured along the
projected $z$-axis), corresponding to $3'.8\to 4'.4$ at the
distance of Mira. An inspection of Figure 1 shows
that this distance approximately corresponds to the
position of condensation A.

The morphologies seen in some of the time frames of Figure 4
are remarkably similar to the head of Mira's
comet. For example, the $t=6.5$, $7.5$ and $9.0\times 10^4$~yr
frames show a ``double bow'' head, in which one could associate
one of the bow shaped structures with condensation A, and the
other one with condensation B (see Figure 1).

In most of the time-frames shown in Figure 4,
we see high density condensations
in the region downstream (i. e., on top) of the star. These
condensations could correspond to some of the knots labeled
E in the schematic diagram of the observations of Mira's
tail (Figure 1).

The model which we are presenting produces a tail with a width of
$\sim 5 \times 10^{17}$~cm (in the $\phi=30^\circ$ column density
map, see Figure 3). This width is similar to the $\sim 7'.5$
($7.2\times 10^{17}$~cm) width of Mira's cometary tail (see Figure 1
of Martin et al. 2007).

\section{Conclusions}

We have presented a 3D simulation of the interaction of a tilted,
latitude-dependent wind with a streaming environment. This simulation
(viewed from an appropriate direction) produces column density maps
which resemble the cometary structure around Mira in a qualitative way.

\begin{figure}
\epsscale{1.0}
\plotone{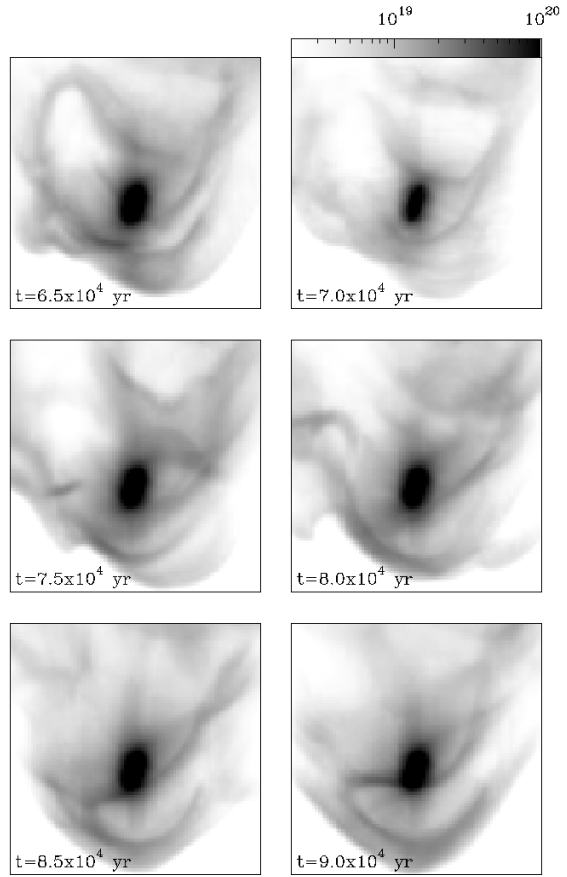}
\caption{Time sequence of column density maps of the head of
the cometary structure (the times in years are given on the bottom
left of each of the frames). The column densities were computed
for the $\phi=30^\circ$ orientation (between the $z$-axis
and the plane of the sky, see the text and Figure 3) of the
motion of Mira. The square domains which are shown have a size
of $10^{18}$~cm, which approximately coincides with the
horizontal extent of the diagram shown in Figure 1.
The column densities (in cm$^{-2}$)
are shown with the logarithmic greyscale given by the bar
on the top right.}
\label{fig4}
\end{figure}

Our simulation produces an asymmetric leading bow shock with a
structure of condensations, and also a complex, time-dependent
tail (see Figure 2). If one compares the predicted column densities
with the UV maps of Martin et al. (2007),
one can see that the individual time-frames (Figure 4)
have condensations that fall at the approximate positions of the
observed condensations (see Figure 1 and Martin et al. 2007). In
particular, our model is successful at reproducing the ``double
bow shock'' structure seen in the head of Mira's comet, with two
bow shaped condensations at approximately the correct distances
away from the star.

This comparison between the predictions and the observations
hinges on the the assumption that the emission coefficient of the observed
emission is approximately proportional to the density, so that the features
seen in the column density maps can indeed be identified with the observed,
emitting condensations. This assumption is of course likely to be
incorrect.

In order to advance in understanding Mira's comet it will be essential
to obtain information about the mechanisms responsible
for the UV emission. Low resolution spectroscopic observations
will provide information about
which mechanisms produce the emission of different regions of
the cometary structure. High resolution observations will provide
radial velocity information, and
proper motion measurements of the condensations of the cometary
structure would also be possible by reimaging
this object within the next few years.

The parameters of our present model are not very well constrained,
except for the known spatial velocity and the orientation of Mira's
motion with respect to the surrounding medium. Also, if one assumes
that the polar axis of the latitude-dependent wind is perpendicular
to the orbital axis of the Mira binary system, one obtains an
observational constraint for the orientation of the polar axis.
We have chosen values for the mass loss rate and the wind velocity
(and for the angular dependence of both of these) which are reasonable
for an AGB star, and a value for the environmental density such as
to give the right size for the head and tail of the cometary structure.

It is notable that through a very limited exploration (with
$\sim 10$ attempts) of the choices for all of these parameters, one
is able to obtain a model that reproduces the remarkable ``double
bow shock'' structure of Mira's comet. We therefore conclude that
the observed double bow shock can be straightforwardly explained
by a model of the interaction of a steady, latitude-dependent wind
with a streaming environment, and that it is not necessary to invoke
the presence of a strong time dependence in the wind in order to
reproduce the observed structures. Wareing et al. (2007) arrived
at a similar conclusion from their analysis of the structure of the
tail of Mira's comet.

We end our discussion by mentioning the differences between our model
and the model presented by Wareing et al. (2007). The main difference
between the two models is that while Wareing et al. (2007) considered
an isotropic wind, we have considered the case of a latitude-dependent
wind. Another, less important difference between the two calculations
is in the way the simulations were initialized.

Wareing et al. (2007)
started their simulation by imposing a wind within a small, spherical
region, having the rest of the computational domain filled in with
a homogeneous, streaming environment. This initialization can be seen
as an idealization of having a star with a wind that suddenly
``turns on'' as it enters its AGB phase, but it is probably not
appropriate for the case of Mira (which was probably in the
AGB phase before starting to interact with the ISM in the galactic
plane).

We have started our simulation assuming that the stellar wind
fills the computational domain, and that the streaming environment
begins to enter the computational domain from one of the grid boundaries.
This is an idealization of the situation in which a stellar wind source
that moves through a very low density environment suddenly enters
a much higher density medium (i. e., the star meets the galactic
plane), which is also an idealized situation that probably does
not correspond to the evolution of Mira's comet (which has been
travelling through the stratified, galactic plane ISM for a much
longer time than the time-integration of our simulation).

However, we find
that for integration times $>5\times 10^4$~yr the structures obtained
in the head of the cometary structure (see, e.g., Figure 4) are
qualitatively similar with both of the two initializations
described above. A detailed description of the long tail of the
cometary tail, however, might require a model in which the
passage of Mira through a more realistic galactic plane ISM is
considered.

\acknowledgements
This work was supported by the CONACyT
grants 46828-F and 61547, the DGAPA (UNAM) grant IN~108207
and the ``Macroproyecto
de Tecnolog\'\i as para la Universidad de la Informaci\'on y la
Computaci\'on'' (Secretar\'\i a de Desarrollo Institucional de la UNAM,
Programa Transdisciplinario en Investigaci\'on y Desarrollo
para Facultades y Escuelas, Unidad de Apoyo a la Investigaci\'on en
Facultades y Escuelas). FDC acknowledges support of the European
Community's Marie Curie Actions - Human Resource and Mobility within
the JETSET (Jet Simulations, Experiments and Theory) network under
contract MRTN-CT-2004 005592.
We thank Enrique Palacios, Mart\'\i n Cruz and Antonio Ram\'\i rez
for supporting the servers in which the calculations of this paper
were carried out. We also thank an anonymous referee for helpful
comments.


\end{document}